\documentclass[superscriptaddress,letterpaper,amssymb,preprint,amsmath,titlepage]{revtex4-1}

\usepackage[english]{babel}
\usepackage[utf8]{inputenc}
\usepackage{mathtools}
\usepackage{mathrsfs}
\usepackage{braket}
\usepackage{xcolor}
\usepackage[symbol]{footmisc}
\usepackage{amsmath}
\usepackage{ragged2e}
\usepackage{xcolor, soul}
\usepackage{lipsum}
\usepackage{siunitx}
\usepackage[labelfont=bf]{caption}
\DeclareCaptionJustification{justified}{\justifying}
\usepackage{subcaption}

\addto\captionsenglish{}
\usepackage{scalerel}
\usepackage{tikz}
\usetikzlibrary{svg.path}
\definecolor{orcidlogocol}{HTML}{A6CE39}
\tikzset{
  orcidlogo/.pic={
    \fill[orcidlogocol] svg{M256,128c0,70.7-57.3,128-128,128C57.3,256,0,198.7,0,128C0,57.3,57.3,0,128,0C198.7,0,256,57.3,256,128z};
    \fill[white] svg{M86.3,186.2H70.9V79.1h15.4v48.4V186.2z}
                 svg{M108.9,79.1h41.6c39.6,0,57,28.3,57,53.6c0,27.5-21.5,53.6-56.8,53.6h-41.8V79.1z M124.3,172.4h24.5c34.9,0,42.9-26.5,42.9-39.7c0-21.5-13.7-39.7-43.7-39.7h-23.7V172.4z}
                 svg{M88.7,56.8c0,5.5-4.5,10.1-10.1,10.1c-5.6,0-10.1-4.6-10.1-10.1c0-5.6,4.5-10.1,10.1-10.1C84.2,46.7,88.7,51.3,88.7,56.8z};}}
\newcommand\orcidicon[1]{\href{https://orcid.org/#1}{\mbox{\scalerel*{
\begin{tikzpicture}[yscale=-1,transform shape]
\pic{orcidlogo};
\end{tikzpicture}
}{|}}}}
\usepackage{xr-hyper}
\usepackage{hyperref} 
\hypersetup{colorlinks,citecolor=red,urlcolor=blue,hypertexnames=true}
\usepackage[capitalise]{cleveref}
\usepackage{filecontents}

\begin{document}
\makeatletter\@input{xSIaux.tex}\makeatother

\title{Investigation of Real-Space Transfer Noise in InP Quantum Wells}

\author{Jiayin Zhang \orcidicon{0009-0005-0736-3554}}
\affiliation{Division of Engineering and Applied Science, California Institute of Technology, Pasadena, CA, USA}

\author{Anthony J. Ardizzi \orcidicon{0000-0001-8667-1208}}
\affiliation{Division of Engineering and Applied Science, California Institute of Technology, Pasadena, CA, USA}

\author{Kieran A. Cleary \orcidicon{0000-0002-8214-8265}}
\affiliation{Division of Physics, Mathematics, and Astronomy, California Institute of Technology, Pasadena, CA 91125, USA}

\author{Austin J. Minnich \orcidicon{0000-0002-9671-9540}
\footnote[1]{Corresponding author: \href{mailto:aminnich@caltech.edu}{aminnich@caltech.edu}}}
\affiliation{Division of Engineering and Applied Science, California Institute of Technology, Pasadena, CA, USA}

\date{\today} 
\begin{abstract}
Indium phosphide (InP) high electron-mobility transistors (HEMTs) are widely used in many fields such as quantum computing because of their unparalleled microwave noise performance. Achieving improved noise performance requires a physical understanding of the noise mechanisms. Here, we experimentally test a theoretical proposal for drain (output) noise as originating in part from real-space transfer (RST) by characterizing the microwave noise temperature of transfer-length method structures with the same channel composition but two different barrier compositions. This choice was made to alter the confining potential of electrons in the channel, thereby affecting the RST mechanism, while avoiding changes to the channel transport properties. We observe trends of noise temperature with physical temperature and source-drain voltage which are compatible with the predictions of RST noise theory. This finding supports the hypothesis that RST contributes to drain noise in HEMTs. 
\end{abstract}


\maketitle

\section{Introduction}

Low noise amplifiers (LNAs) based on InP HEMTs are widely used in fields such as radio astronomy, weather and climate studies, and quantum computing due to their excellent noise performance at room and cryogenic temperatures \cite{bardin_microwaves_2021, pospieszalski_extremely_2005,krantz_quantum_2019}. Indium phosphide (InP) HEMTs currently exhibit the best noise performance among all HEMT techonologies. For state-of-the-art InP HEMTs, the reported cryogenic noise figures are around 5 times the standard quantum limit in the 1 - 100 GHz frequency range \cite{schleeh_characterization_2013,cuadrado-calle_broadband_2017,jo_l_2020,cha_inp_2020,heinz_50-nm_2021,cha_optimization_2023,li_investigation_2024}. 

To further improve noise performance, understanding the physical origin of the noise is of high importance. According to the Pospieszalski model, noise in HEMTs is attributed to gate noise and drain noise at the input and output of the device, respectively \cite{pospieszalski_modeling_1989}. It is generally accepted that gate noise originates from Nyquist-Johnson noise in the gate metal at the gate physical temperature \cite{pospieszalski_dependence_2017,schleeh_phonon_2015, ardizzi_anthony_j_self-heating_2022}. Drain noise, on the other hand, has traditionally been used as a fitting parameter without an accepted physical explanation. One of the earliest theories by Pucel et al. attributed drain noise to the generation of dipole layers \cite{pucel_signal_1975}. Several studies have proposed that drain noise arises from suppressed shot noise \cite{gonzalez_noise_1997,pospieszalski_limits_2017}, while others suggested the origin as impact ionization \cite{shigekawa_impact-ionization-induced_1999,ruiz_impact_2019,yuce_large-signal_2024}.

A recent theoretical study by Esho et al. attributed drain noise to the real-space transfer (RST) mechanism \cite{esho_theory_2022}. Real-space transfer is a process in which channel electrons are thermionically emitted over the confining potential at the heterointerface into the adjacent barrier film, thereby transferring their location in real space. It is distinguished from intervalley scattering in which electrons change their state in reciprocal space rather than real space\cite{price_intervalley_1960}.  Real-space transfer is possible in HEMTs because electric fields may exceed 20 kV/cm under the gate, heating the electrons out of equilibrium with the lattice to temperatures exceeding 1000 K. \cite{schwierz_modern_2003, shah_investigation_1981} Real-space transfer was originally proposed as a means to achieve negative differential resistance via transport parallel to an interface, \cite{hess_negative_1979} and various transferred-electron devices have been proposed based on this effect \cite{kastalsky_novel_1983,luryi_charge_1984}. Microwave partition noise arises from real-space transfer because the electrical transport properties of the channel and barrier films are very different, satisfying the requirements of the partition noise mechanism \cite{ambrozy_electronic_1982}. 



The investigation of RST as contributing to microwave noise of quantum wells has a long history. Aninkevičius et al. attributed the hot electron noise in the AlGaAs/GaAs heterostructure to RST\cite{aninkevicius_comparative_1993}. Matulionis et al. experimentally studied microwave noise in InAlAs/InGaAs/InAlAs quantum wells, confirming that there exists a strong dependence of drain noise on the electric field that was attributed to RST \cite{matulionis_qw-shape-dependent_1997}. It was also reported that increasing the conduction band offset in the InP HEMT heterojunction improves quantum confinement and thus suppress RST noise \cite{matulionis_hot-electron_1999}. Recently, Gabritchidze et al. characterized GaAs mHEMTs and provided evidence that RST contributes to drain noise in HEMT devices \cite{gabritchidze_experimental_2022}. Li et al. also attributed the excess noise part of the drain noise to RST noise in state-of-the-art InP HEMTs \cite{li_epitaxial_2024}. However, additional direct evidence for the contribution of RST to microwave noise in InP quantum wells used in modern HEMTs, without the complication of the additional noise sources present in full HEMTs, is important to further validate the RST theory.






Here, we report an experimental study of microwave noise in transfer-length method (TLM) structures fabricated on state-of-art epitaxial quantum wells used for InP HEMTs. By characterizing gate-less devices, the contribution of the gate noise is eliminated. Two sets of ungated HEMT devices are fabricated on two InP substrates, with the only difference being the barrier indium composition. This approach allows the confining potential of the electrons in the channel to be modified, thereby affecting the RST mechanism, while avoiding any change to the transport properties of the channel electrons. The measured trends are in good qualitative agreement with a model which attributes noise to a combination of thermal and RST noise. This finding suggests that drain noise in InP HEMTs arises from the same mechanisms and that RST noise could be suppressed by improving quantum confinement of channel electrons.




\section{Theory} \label{sec:theory}
We begin by presenting the equations for noise temperature that will be used to interpret the experimental measurements.  Our model for microwave noise in the TLM structures attributes the noise to two components: noise from the RST mechanisms with a noise temperature $T_{RST}$, and thermal noise from the electron gas at the physical electron temperature $T_\text{e}$. We first consider the noise contributed by RST as derived by Esho et al. \cite{esho_theory_2022}. The RST partition velocity fluctuation noise power spectral density $S_{\nu,RST}$ can then be written as (see \cite{hartnagel_microwave_2001} Eq. 9.6) as:
\begin{equation}
    \label{eq:RST2}
    S_{\nu,RST} = 4\frac{n_{ch}n_{ba}}{n^2}(v_{ch}-v_{ba})^2\frac{\overline{\tau}_{RST}}{1+(\omega\overline{\tau}_{RST})^2}
\end{equation}
where $n_{ch}$ ($n_{ba}$) is the carrier density in the channel (barrier),  $v_{ch}$ ($v_{ba}$) is the drift velocity in the channel (barrier), $n=n_{ch}+n_{ba}$ is the total carrier density, $\omega=2\pi f$ is the angular frequency, and $\overline{\tau}_{RST}$ is the RST relaxation time.



The Ramo-Shockley theorem can be used to relate the velocity fluctuation and current noise power spectral densities (see \cite{hartnagel_microwave_2001} Eq 8.4):
\begin{equation}
    \label{eq:RST3}
    S_I = \frac{e^2nW}{L}S_\nu
\end{equation}
where $e$ is the electron charge, and $W$ and $L$ are the width and length of the channel, respectively.

Since emission across the channel-barrier interface is thermionic, the barrier electron density $n_{ba}$ depends on the channel electron density $n_{ch}$ (see \cite{sze_physics_2007} Ch. 3 Eq. 61) as:
\begin{equation}
    \label{eq:RST4}
    n_{ba} = \gamma\eta n_{ch}
\end{equation}
where $\gamma$ is the probability for a hot electron to emit across the barrier, $\eta=\exp{[-\Delta E_c/(k_BT_e)]}$, is the fraction of electrons with sufficient thermal energy to cross the interface, $T_e$ is the electron temperature, and $\Delta E_c$ is the conduction band offset between InGaAs and In$_x$Al$_{1-x}$As. For simplicity, we assume $\gamma=1$ in this study.

The effective noise temperature for an arbitrary current noise power spectral density in the direction parallel to an applied electric field $E$ can be defined using the generalized Johnson-Nyquist equation as:
\begin{equation}
    \label{eq:RST5}
    T_{N,||}(E) = \frac{S_{I||}(E)\Re(Z_{||}(E))}{4k_B}
\end{equation}
where $k_{B}$ is the Boltzmann constant, and $\Re(Z_{||}(E))$ is the real part of the electrical impedance along the direction of the applied field \cite{hartnagel_microwave_2001}. 

Finally, we consider the limit $\omega\overline{\tau}_{RST}\ll1$, which is justified in the microwave regime of approximately $f<100$~GHz because $\overline{\tau}_{RST} \sim 10$ ps, \cite{hartnagel_microwave_2001} and $v\equiv v_{ch}\gg v_{ba}$ due to the much higher mobility of the channel compared to the barrier. Combining the above equations, we obtain the RST noise temperature ($T_{RST}$) as: 

\begin{equation}
    \label{eq:TRST}
    T_{RST}(E) = \frac{e^{2}nv(E)^{2}\overline{\tau}_{RST}}{k_{B}}\frac{W_\text{re}}{L_\text{re}}\frac{\eta(E)}{(1+\eta(E))^{2}}\Re(Z(\omega,E))
\end{equation}
where $W_\text{re}$ and $L_\text{re}$ are the width and length of the recess, respectively.

For the thermal noise of the channel electrons, an expression for $T_\text{e}$ is obtained using the standard formula for Joule heating of electrons with electric field $E$ and assuming electrons lose energy to the lattice with an energy relaxation time $\tau_E$ \cite{matulionis_qw-shape-dependent_1997}:
\begin{equation}
    \label{eq:Te}
    T_{e}(E) = T_{lat}+\frac{e\mu(E)\overline{\tau}_{E}}{k_{B}}E^{2}
\end{equation}
where $T_{lat}$ is the lattice temperature which is assumed to be the ambient temperature, $\mu$ is the channel electron mobility, and $\overline{\tau}_{E}$ is the electron energy relaxation time. 


The final expression for the drain noise along the direction of the applied field ($T_{N||}$) can then be written as:

\begin{equation}
    \label{eq:TN}
    T_{N||}(E) = T_{RST}(E)+T_{e}(E)
\end{equation}

because the thermal and RST noise sources are uncorrelated. For fitting purposes, the hot-electron energy relaxation time $\overline{\tau}_{E}$ and the RST relexation time $\overline{\tau}_{RST}$ are the only free fitting parameters.

\section{Experiment}

\subsection{Sample fabrication}\label{sec:Fabrication}

\begin{figure}[t]
    {\label{fig:structure}}
    {\phantomsubcaption\label{fig:DeviceStructure}}
    {\phantomsubcaption\label{fig:DeviceSEM}}
    \centering
    \includegraphics[width=1.0\textwidth]{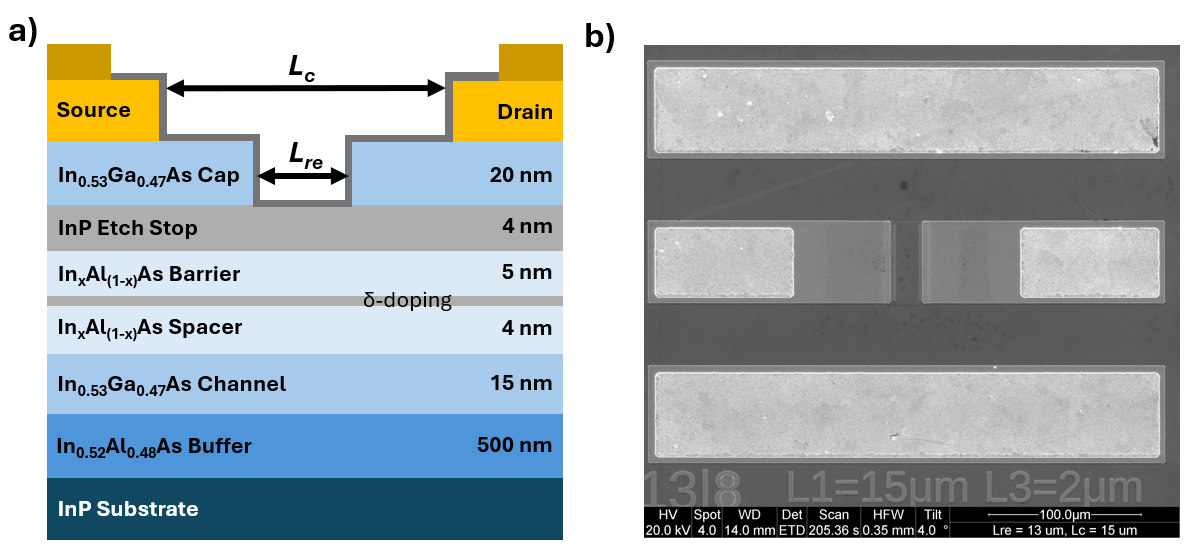}
    \caption{\textbf{(a)} Cross section schematic of the ungated HEMT device. The epitaxial layers are grown on an InP substrate. From bottom to top, the epitaxial layers include a 500 nm In$_\text{0.52}$Al$_\text{0.48}$As  buffer, a 15 nm In$_\text{0.53}$Ga$_\text{0.47}$As channel, a 9 nm In$_\text{x}$Al$_\text{1-x}$As barrier/spacer, a 4 nm InP etch stop, and a 20 nm In$_\text{0.53}$Ga$_\text{0.47}$As cap. The $\delta$ doping is located at 4 nm above the channel. A recess with controlled recess length $L_\text{re}$ is etched on the cap to the etch stop. \textbf{(b)} Top view SEM image of the ungated HEMT device. The cross section shown in (a) is located at the center of the device. }
\end{figure}

TLM structures were fabricated on epitaxial films with the structure shown in \cref{fig:DeviceStructure}. The wafers were grown by molecular beam epitaxy by IntelliEpi. Two sets of samples were used in which the substrates were identical except for the barrier (In$_x$Al$_{1-x}$As) Indium composition, $x$. For the lattice-matched sample (LM), $x = 0.52$, and for the tensile-barrier sample (TB), $x = 0.45$. With these barrier compositions, the conduction band offsets ($\Delta E_{c}$) of the LM sample and TB samples were calculated as 0.52 eV and 0.66 eV, respectively, using conduction band offsets and bowing parameter values from Vurgaftman et al \cite{vurgaftman_band_2001}. As a result, the channel transport properties of each sample are nominally the same, but the electrons have greater quantum confinement in the TB sample.

The substrates were diced into $1 \times 1$ cm\textsuperscript{2} chips for further processing. Each TLM structure was designed as a co-planar wave guide (CPW). \Cref{fig:DeviceSEM} shows a top view scanning electron microscope (SEM) image of the device. This design enabled the devices to be mounted in a home-built cryogenic probe station for three-pin ground-source-ground (GSG) probing \cite{russell_cryogenic_2012}. The samples were fabricated with various recess lengths ($L_\text{re}$) of 1, 3, 8, 13, and 18 \si{\um}. The side wall of the recess was fixed to be 1~\si{\um} away from the edge of the source and drain metal, such that the distance between the Ohmic contacts ($L_{c}$) was  3, 5, 10, 15, and 20 \si{\um}, respectively. 

Two other sets of devices were fabricated to determine the contact resistance using the TLM method in \cite{lamarre_ohmic_1993,lee_theoretical_2021}. One set of devices had fixed $L_\text{re} = 1$ \si{\um} and varying $L_{c} = $ 3, 5, 10, 15, and 20 \si{\um}, while the other set had no recess ($L_{re}=0$) but the same variation in $L_{c}$. The latter set was used to determine the contact resistance and sheet resistivity of the cap layer. With these three sets of devices, contact resistance, and sheet resistances of the cap and channel, as well as the effective channel-barrier resistivity, were determined. These sets of devices were then patterned periodically across the $1 \times 1$ cm\textsuperscript{2} chip.

For fabrication of devices for microwave noise characterization, mesas were first created by wet etching. A citric acid and hydrogen peroxide mixture was used to selectively etch the InGaAs cap layer, hydrochloric acid was used to selectively etch the InP etch stop, and a phosphoric acid and hydrogen peroxide mixture was used to etch into the buffer. Then an Ohmic contact metal stack of Pt/Ti/Pt/Au (9/15/15/20 nm) at the source and drain was deposited by electron-beam evaporation. The room-temperature contact resistance to the cap was measured to be 40 m$\Omega$ mm. The recess was patterned by electron beam lithography and was etched to the InP etch stop using a citric acid/hydrogen peroxide mixture. Next, the overlay metallization layer of Ti/Au (10/300 nm) was deposited by electron beam evaporation, and the devices were passivated with 30 nm of Al$_\text{2}$O$_\text{3}$ using thermal ALD at 300 \si{\celsius}. Finally, the overlay metallization pads were plated with $\sim3$ \si{\um} of Au using electroplating for probing.

\subsection{Microwave characterization}\label{sec:Measurement}

We measured the microwave S-parameters and noise temperature with 50~$\Omega$ source impedance ($T_{50}$) in a custom-built probe station. The probe station has been described previously \cite{russell_cryogenic_2012}, with some modifications made for this work. In particular, the modifications enabled noise measurements of lossy two-port devices such as the TLM structures measured here, in contrast to active HEMTs with gain; as well as pulsed measurements to enable sufficiently high biases to observe RST noise while minimizing excessive self-heating of the samples.


Full details of the calibration, measurement, and data processing procedure are available in \cite{ardizzi_variable_2025}. In brief, pulsed biasing was performed using an arbitrary waveform generator, which supplied voltage pulses with a 20~ms period and 500~$\mu$s pulse width, for a duty cycle of 2.5\%. Voltage and current time series were measured simultaneously with both S-parameter and noise measurements on an oscilloscope using a custom high-speed current probe. The pulse parameters were chosen to avoid excessive Joule heating of the channel, which has been shown to mask and in some cases suppress RST \cite{hartnagel_microwave_2001}. The pulse duration was limited by the time-domain resolution of the vector network analyzer (VNA) used for S-parameter measurements. In general, the DUT gain can be extracted from Y-factor measurements, but since the resistance of the ungated structures changes appreciably over the range of biases measured here, the impedance match becomes poor at high biases and must be considered in the analysis.



The microwave noise temperature ($T_{50}$) in 4-8 GHz was measured using the Y-factor method \cite{arthur_measurement_1974}. Noise power at the output of the probe station was measured using a home-built backend receiver with an InP-based low-noise amplifier (Low Noise Factory LNF-LNC4\_8C) as the first stage amplifier. A mixer was used to down-convert from 5~GHz into an 800~MHz wide band centered at DC, with a 27.5~MHz high-pass filter employed to avoid 1/f noise. A zero-bias tunnel-diode detector with ultra-fast response time of $\sim5$~ns was used to linearly convert RF noise power into a DC voltage. A time series measurement was then performed using an oscilloscope, synchronized with the pulse biasing and IV-curve oscilloscope measurement.

The backend noise temperature referred to the input of the output probe tip was measured using the Y-factor method, with a room-temperature and liquid nitrogen cooled 50~$\Omega$ load at the probe station input as the hot and cold reference loads, respectively. The value was measured as $T_\text{BE}=132\pm9$~K. Corrections were applied to account for the S-parameters and physical temperature of the coaxial cabling from the probe station input to the probe tip (See SI of Ref. \cite{ardizzi_self-heating_2022}). 

A high excess-noise-ratio noise source (ENR$\sim25$~dB at 5~GHz; model MMW TECHNOLOGY INC 10MHz--40GHz NOISE SOURCE S/M12) was used to supply input noise. The Y-factor method, along with the previously measured $T_\text{BE}$, was used to calibrate the noise temperature at the output of the input probe tip. These were measured to be $T_{h}=40916\pm822$~K and $T_{c}=295.2\pm0.9$~K at 5~GHz with the noise source on and off, respectively. This temperature range enabled accurate noise measurements at high biases where the DUT noise is in excess of 10,000~K and the loss approaches $20$~dB, while sacrificing measurement precision at low biases where the noise is on the order of room temperature.
The S-Parameter and Y-factor noise power of a series of TLM structures of differing gate recess lengths were then measured over a range of biases, and $T_{50}$ was extracted using the measurements of $T_\text{BE}$, $T_{h}$ and $T_{c}$. In order to extract the effective noise temperature of the recessed region of the device, impedance mismatch corrections were applied to extract the noise temperature of the lumped TLM device (a full description of this correction is given in Section II of \cite{ardizzi_variable_2025}, see specifically equation 11). The noise contributions from the access region, particularly the ohmic contacts and unetched cap, were then assumed to be negligible compared to the noise originating in the recessed region since the access region noise does not depend on bias and can contribute at most thermal noise.


\section{Results}

\subsection{DC and RF characteristics}\label{sec:Data}

\begin{figure}[t]
    {\phantomsubcaption\label{fig:IV}}
    {\phantomsubcaption\label{fig:Spar}}
    \centering
    \includegraphics[width=1.0\textwidth]{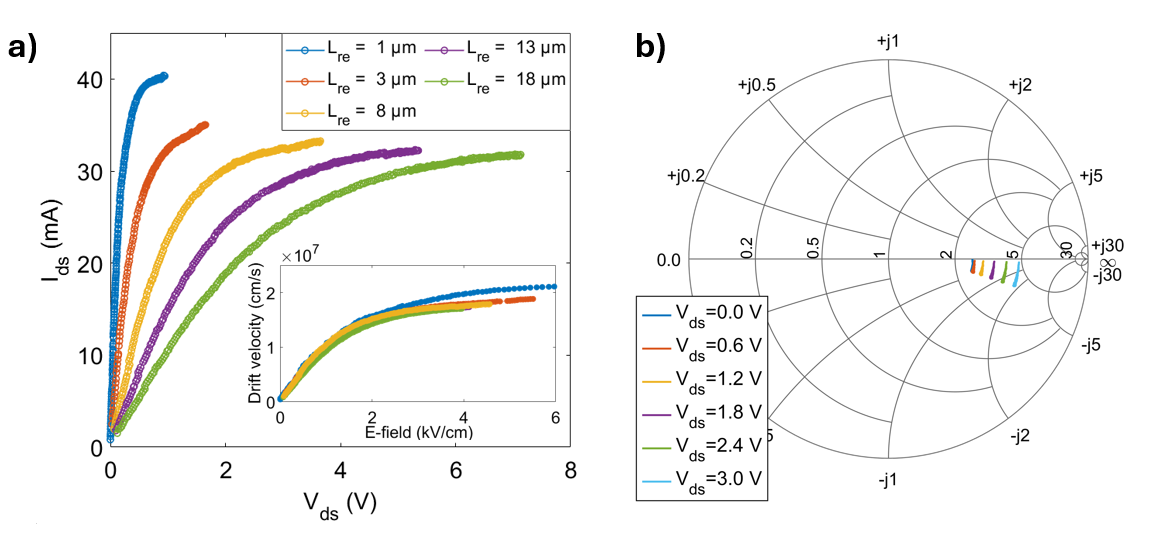}
    \caption{\textbf{(a)} Room-temperature I-V characteristics of the LM device. Inset: drift velocity versus electric field. \textbf{(b)} Representative $S_{11}$ data on a LM device with $L_\text{re}$ = 13 \si{\um} from 3.5 - 8.5 GHz.}
\end{figure}

We begin by showing representative I-V characteristics for the LM device in \Cref{fig:IV}. In the low-bias linear regions of the I-V curves, longer gate recesses produce IV curves with smaller slopes due to the larger total resistance of the longer channel. All I-V curves eventually saturate at high bias. From these measurements, electron mobilities ($\mu$) and drift velocities ($v$) can be obtained for further analysis.

The electric field in the channel is determined from the TLM geometry, contact resistance, and applied bias. To do so, the resistance of the bias lines was measured by landing on an open gold pad, yielding the voltage dropped only across the DUT. Then, the total resistance of the device was compared to the statistical average of 5 nominally identical devices, and the difference was attributed to a difference in contact resistance. This offset was then subtracted from the nominal (statistically averaged) contact resistance to find the per-device contact resistance, yielding the voltage $V_{ch}$ dropped across the channel. The electric field was then computed as $E=V_{ch}/L_\text{re}$. While the variation in contact resistance was typically better than 5\%, the above calibration procedure was found to produce the best agreement in drift velocity versus electric field across devices of different channel lengths. As shown in the inset of \Cref{fig:IV}, the drift velocities collapse onto a single curve when plotted versus electric field, as expected.

The microwave S-parameters were also measured as shown in \cref{fig:Spar}. The devices are poorly matched to 50 $\Omega$, especially at higher biases. For example, for a $L_\text{re}$ = 13 \si{\um} device on the LM sample between 3.5 - 8.5 GHz,  $S_{11}$ is approximately 5 $\Omega$. However, this mismatch is accounted for using the methods described in \cite{ardizzi_variable_2025}.

\begin{figure}[t]
    {\phantomsubcaption\label{fig:Yfactor}}
    {\phantomsubcaption\label{fig:ZoomedYfactor}}
    \centering
    \includegraphics[width=1.0\textwidth]{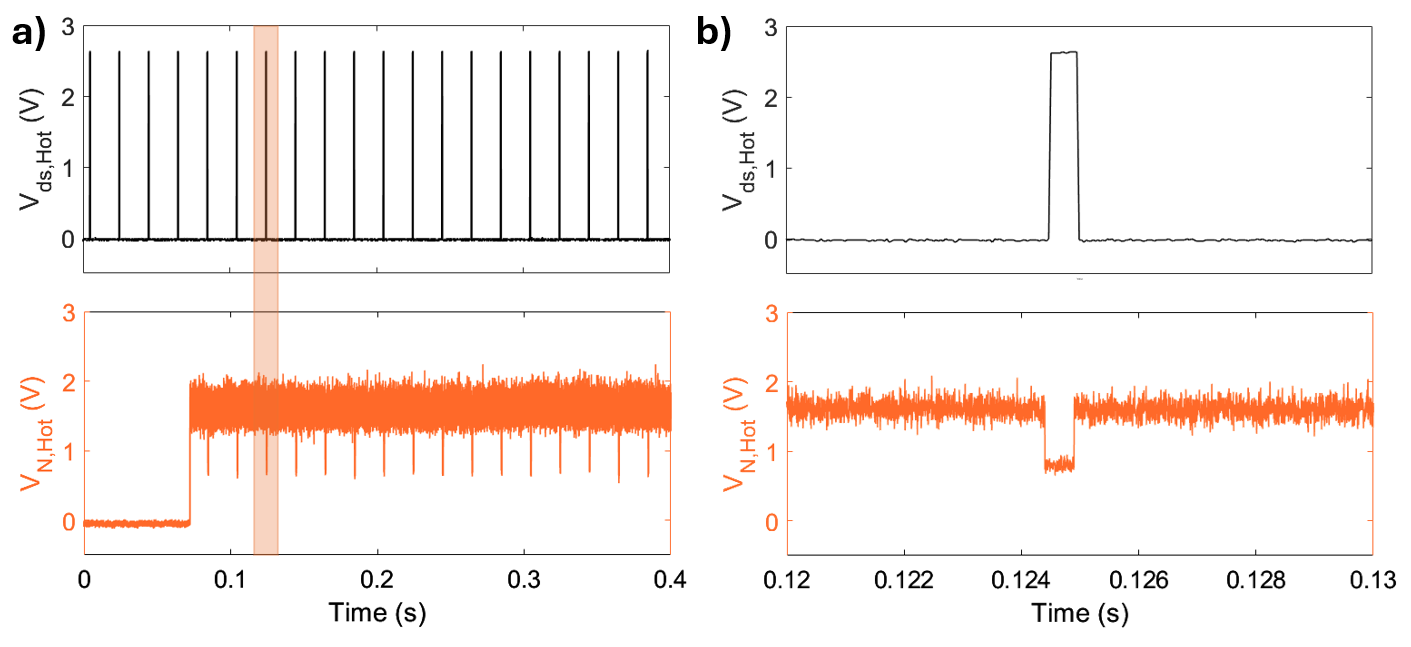}
    \caption{\textbf{(a)} Representative raw noise power data versus time for a LM sample with recess length $L_\text{re}$ = 13 \si{\um} at 5 GHz and room temperature. The pulsed bias $V_\text{ds,Hot}$ (black line) and the diode detector hot voltage $V_\text{N,Hot}$ (orange line) are shown. \textbf{(b)} Zoom-in view of (a) showing a single bias pulse and the noise power data respectively. The offset between the bias pulse and noise measurement is due to a slight difference in trigger timing between the noise voltage and bias voltage oscilloscopes, which is accounted for in software.}
\end{figure}

We next present the raw data from the microwave noise measurement. \Cref{fig:Yfactor} plots a representative hot noise voltage time series data (for simplicity, cold noise voltage data is not shown). The upper panel shows the pulsed bias applied to the device with a  20~ms period and 500~$\mu$s pulse width. The lower panel shows the measured hot noise power with pulsed biasing. The first $\sim0.08$~s shows the zero-power voltage offset ($V_0$) calibration for the tunnel-diode detector.



\Cref{fig:ZoomedYfactor} shows a magnified view of an individual bias pulse and the associated change in noise voltage. In this case, the noise, loss, and impedance mismatch all simultaneously increase with bias, ultimately manifesting as a decrease in measured noise power with bias. The noise voltage measured with the bias on was averaged across all 20 pulses, yielding the average hot noise power arriving at the detector. The Y-factor was then calculated by repeating this procedure with the noise source off and calculating the ratio $Y=(\overline{V}_\text{N,Hot}-\overline{V}_{0,\text{Hot}})/(\overline{V}_\text{N,Cold}-\overline{V}_{0,\text{Cold}})$, where $\overline{V}_\text{N,Hot}$ and $\overline{V}_\text{N,Cold}$ are the averaged hot and cold noise power with the noise source on, respectively, and $\overline{V}_{0,\text{Hot}}$ and $\overline{V}_{0,\text{Cold}}$ are the hot and cold noise power with the noise source off, respectively.

\subsection{Microwave noise temperature }\label{sec:NoiseTemp}

We next present the microwave noise data obtained from the Y-factor measurements. The lumped noise temperature of the DUT is extracted from the measured Y-factor. This lumped noise consists of the noise from the access regions and the intrinsic channel region. Since the noise from the access regions for all devices does not vary with bias, the bias-dependence of the lumped noise temperature coincides with that of the intrinsic channel region.



\begin{figure}[t]
    {\phantomsubcaption\label{fig:TNvsLre_LM}}
    {\phantomsubcaption\label{fig:TNvsLre_TB}}
    \centering
    \includegraphics[width=1.0\textwidth]{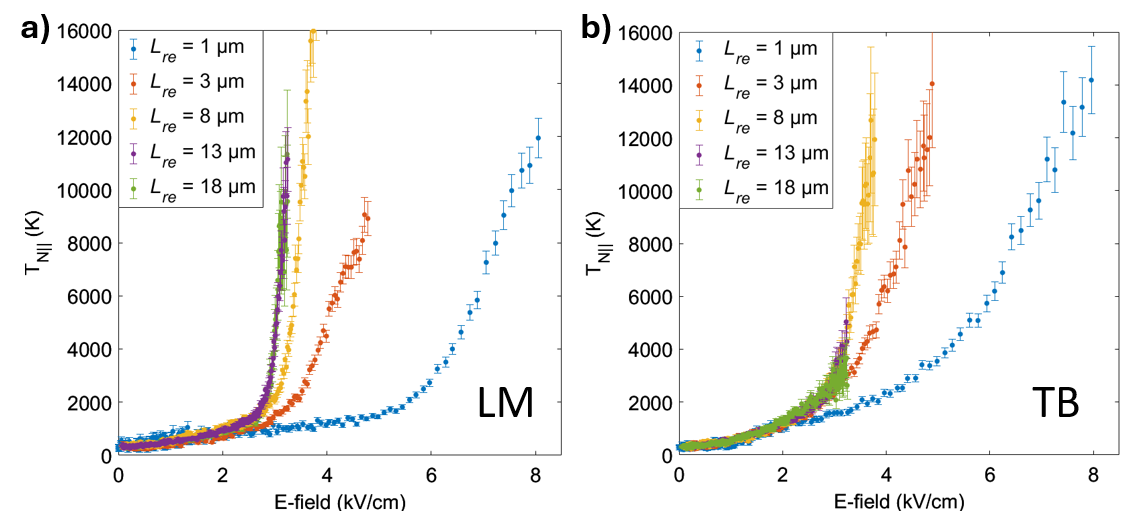}
    \caption{\textbf{(a)} Noise temperature parallel to the channel ($T_{N}||$) of LM sample vs electric field. All data are taken at 5 GHz at room temperature. \textbf{(b)} Analogous plot for TB sample.}
    {\label{fig:TNvsLre}}
\end{figure}

\Cref{fig:TNvsLre_LM} plots the measured noise temperature of devices of all recess lengths on the LM sample. All devices show an initial increase in noise with field, followed by a rapid increase in noise at an electric field value that depends on the recess length, $L_{re}$.  For the $L_\text{re}$ = 18 \si{\um} device, this field is around 2.5 kV/cm, while for the $L_\text{re}$ = 1 \si{\um} device, this increase is not observed until approximately 5.3 kV/cm. This recess-length-dependent trend can also be observed on the devices fabricated on the TB chip, as shown in \Cref{fig:TNvsLre_TB}. From these measurements, it is observed that with a shorter recess length, the rapid increase in noise temperature  occurs at a higher electric field.

The qualitative features in \Cref{fig:TNvsLre} are compatible with expectations from RST noise theory. As discussed in \cite{hartnagel_microwave_2001}, a length-dependence for the onset of the rapid increase in noise can be attributed to the following mechanism. For longer recesses, the transit time from  source to drain is longer, increasing the probability that channel electrons will thermionically emit into the barrier. This point implies that even at the same electric field, the noise for a device with a longer recess will be higher because of the higher probability of real-space transfer. An additional conclusion is that the noise trend measured in this study is unlikely to be caused by impact ionization, which is not expected to exhibit a length dependence of noise for a given electric field.


\subsection{Comparison to noise model}\label{sec:RSTModel fitting}
\begin{figure}[t]
    \centering
    \includegraphics[width=1.0\textwidth]{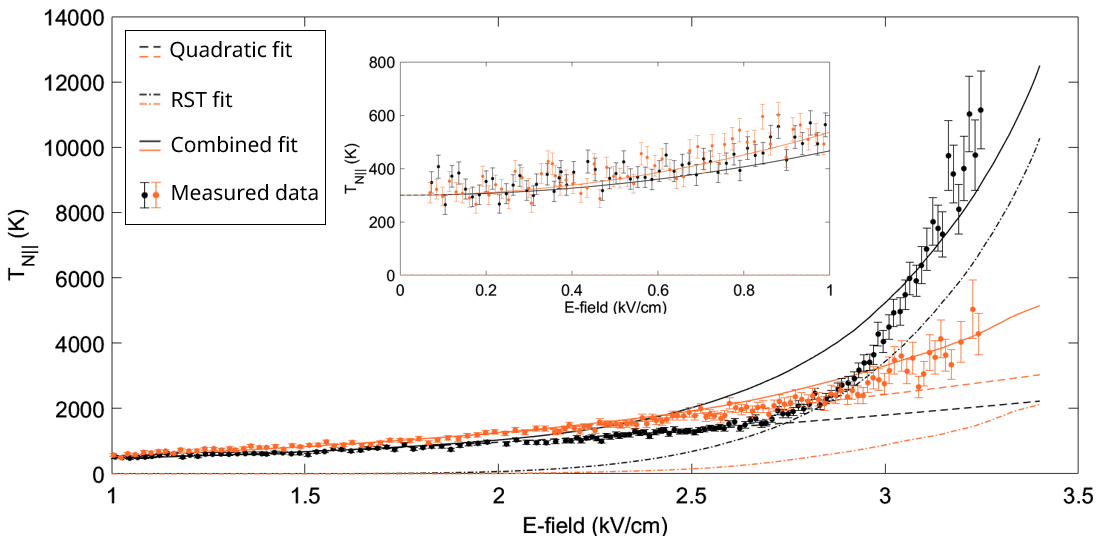}
    \caption{Noise temperature parallel to the channel ($T_{N}||$) of LM and TB samples versus electric field. The black (orange) data are taken on  $L_\text{re}$ = 13 \si{\um} devices on the LM (TB) sample at 5 GHz at room temperature. The dashed lines are the predicted electron temperature ($T_{e}$) from \cref{eq:Te}. The dash-dotted lines are the calculated real-space transfer noise ($T_{RST}$) from the model described in \cref{eq:TN}. The solid lines are the combined noise temperature predicted by the electron temperature and the real-space transfer model. Inset: low field region, showing that both fitted model and the measured data are about 300 K at zero field.}
    {\label{fig:TNvsEfield_LMTB}}
\end{figure}

We now use the noise model from \Cref{sec:theory} to interpret the experimental data. \Cref{fig:TNvsEfield_LMTB} presents the noise temperature ($T_{N||}$) of both LM and TB samples versus electric field for $L_\text{re}$ = 13 \si{\um} devices on both LM and TB samples at 5 GHz at room temperature shown in \cref{fig:TNvsLre}. The black dashed lines in \cref{fig:TNvsEfield_LMTB} are the predicted electron temperature ($T_{e}$) calculated by \cref{eq:Te}. By fitting \cref{eq:Te} to the measured data in the low-field region, the electron energy relaxation time-constant $\overline{\tau}_{E}$ was obtained. For this pair of devices, $\overline{\tau}_{E}$ is 1.40 ps for the LM sample and 1.96 ps for the TB sample. These values are in semi-quantitative agreement with values reported previously (c.f. \cite{matulionis_hot-electron_1999} Table I). The inset of \Cref{fig:TNvsEfield_LMTB} shows that for both LM and TB samples, at zero field and room temperature, the fitted and measured noise data are approximately 300 K, as expected.




The RST temperature ($T_{RST}$) and the total noise temperature along the direction of the applied field ($T_{N||}$) are  calculated by the model described in \cref{eq:TRST} and \cref{eq:TN}, respectively. In the model, the carrier density $n$ is obtained from Hall measurements, the drift velocity of the electrons $v$ is obtained from the I-V curves shown in \cref{fig:IV}, and the resistance of the channel $\Re(Z(\omega,E))$ is taken from the TLM calibration of the fabricated devices. The RST relaxation time constant $\overline{\tau}_{RST}$ is fitted to the measured data to yield $\overline{\tau}_{RST}=40$~ps (6 ps) for the LM (TB) sample. This is on the order of published literature results on similar structures which estimated RST relaxation times on the order of $\sim50-100$~ps (c.f. Ref. \cite{hartnagel_microwave_2001} Ch. 17.1).


\begin{figure}[t]
    {\phantomsubcaption\label{fig:TvsE_1um}}
    {\phantomsubcaption\label{fig:TvsE_3um}}
    {\phantomsubcaption\label{fig:TvsE_8um}}
    {\phantomsubcaption\label{fig:TvsE_18um}}
    \centering
    \includegraphics[width=1.0\textwidth]{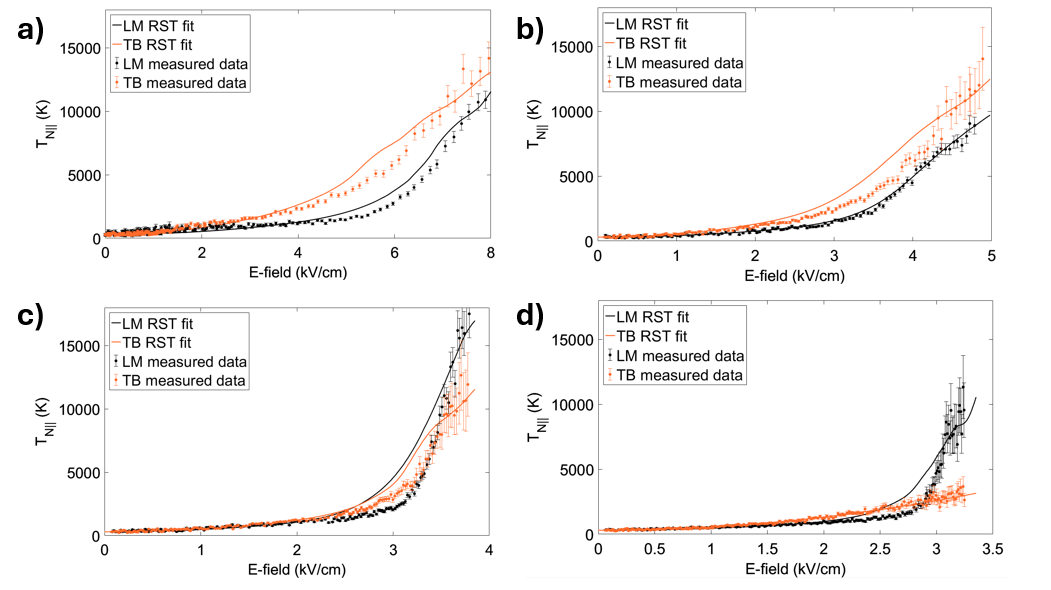}
    \caption{\textbf{(a)} Noise temperature parallel to the channel ($T_{N}||$) of LM sample (black data) and TB sample (orange data) vs electric field. The data are taken on  $L_\text{re}$ = 1 \si{\um} devices at 5 GHz at room temperature. The black line is the calculated noise from the model of the LM sample. The orange line is the calculated noise from the model of the TB sample.\textbf{(b)} Respective data taken and calculated on $L_\text{re}$ = 3 \si{\um} devices. \textbf{(c)} Respective data taken and calculated on $L_\text{re}$ = 8 \si{\um} devices. \textbf{(d)} Respective data taken and calculated on $L_\text{re}$ = 18 \si{\um} devices.}
    {\label{fig:TN_All}}
\end{figure}

The noise model predicts a quasi-exponential dependence of $T_{RST}$ on electric field, since $T_{RST}$ depends exponentially on $T_{e}$ through $\eta(E)$, and $T_{e}$ depends in turn on $E^2$. As shown in \cref{fig:TNvsEfield_LMTB}, the contribution of $T_{RST}$ is almost zero until the electric field reaches approximately 1.9 kV/cm. At about 2.8 kV/cm, the contribution of $T_{RST}$ surpasses the contribution of $T_{e}$, and dominates the overall noise contributions. The combined solid curve calculated from \cref{eq:TN} qualitatively captures this exponential dependence on noise temperature and matches the measured data.

\Cref{fig:TN_All} plots the measured noise data and the calculated noise temperature of $L_\text{re}$ = 1 \si{\um}, $L_\text{re}$ = 3 \si{\um}, $L_\text{re}$ = 8 \si{\um}, and $L_\text{re}$ = 18 \si{\um} for LM and TB samples. For $L_\text{re}$ = 1 \si{\um} and $L_\text{re}$ = 3 \si{\um} devices, at the largest electric field we measured, the TB devices exhibit higher noise temperature. For $L_\text{re}$ = 8 \si{\um} and $L_\text{re}$ = 18 \si{\um}, the relationship is similar to that of the $L_\text{re}$ = 13 \si{\um} devices presented in \cref{fig:TNvsEfield_LMTB}. These measurements show that the influence of RST is less pronounced for the shorter recess lengths, consistent with the results in \Cref{fig:TNvsLre} and those reported previously \cite{hartnagel_microwave_2001}. Additionally, at low field, the TB samples exhibit higher noise than the LM samples, indicated also by a larger extracted energy relaxation time. 

This dependence of energy relaxation time on strain has been observed in the InGaAs quantum well literature \cite{matulionis_hot-electron_1999}. A possible explanation is as follows: the strain near the channel/barrier interface  tends to increase the energy of optical phonons, which are responsible for the majority of electron-phonon interactions at room temperature. Since the electron-phonon interaction occurs primarily with the hottest electrons (those in the high-energy tail of the hot-electron distribution), the increase in phonon energy leads to an increase in the steady-state electron temperature at which the energy input from the bias is balanced by energy loss to the lattice. This effect has also been attributed to the higher mobility observed in strained-channel devices \cite{malmkvist_epitaxial_2008}.


While this contribution runs opposite to the reduction in RST noise for the TB sample, we emphasize that the noise in the LM sample increases above the noise in the TB sample at sufficiently high electric field, indicating that the RST mechanism ultimately dominates at high fields. This is particularly evident in $L_\text{re}$ = 8 \si{\um}, $L_\text{re}$ = 13 \si{\um}, and $L_\text{re}$ = 18 \si{\um} devices. For devices $L_\text{re}$ = 1 \si{\um} and $L_\text{re}$ = 3 \si{\um}, in our measurement range, which is limited by the error of the Y factor measurement, the TB samples still exhibit a higher noise temperature, but trends indicate that at a higher electric field, the noise temperature of the LM samples will eventually surpass the TB samples as in devices of longer recess lengths. 


We attribute the model overestimation, for example between 2~kV/cm and 3~kV/cm for the $L_\text{re}$ = 13 \si{\um} devices in \cref{fig:TNvsEfield_LMTB}, to two factors. First, the noise model does not consider the effects of convective noise which arise from fluctuations in the carrier mobility, and which can partially suppress current fluctuations in cases where the mobility decreases with increasing energy \cite{skullerud_longitudinal_1969,gantsevich_theory_1979} as is the case in semiconductors. Second, in the region where $T_{N||}$ changes most rapidly with $E$, the model is sensitive to changes in several parameters including the electric field value, the impedance correction, and the extracted $\mu(E)$ and $v(E)$. It is possible that that small errors in these calibrations will influence the quantitative trend from the model. However, it is clear that the model qualitatively captures the key trends in the noise data, providing support for the physical origin of the microwave noise as a combination of thermal and RST noise.

\section{Discussion}


This study has shown that a noise model consisting of thermal and RST noise can account for the experimental trends of TLM devices. In particular, the observations of a rapid increase in noise above a certain electric field, the dependence of that electric field value on recess length, and the varying contribution of RST noise for different recess lengths, are all compatible with the predictions of RST theory. The predicted dependence of the noise on the barrier composition is also observed. This finding agrees with prior studies of microwave noise in quantum wells, for instance by Aninkevičius et al. for AlGaAs/GaAs quantum wells. \cite{aninkevicius_comparative_1993} Our study shows these effects, previously observed in other quantum wells, remain relevant for  state-of-the-art InP HEMT epitaxial materials. The good qualitative agreement with the RST model proposed by Esho et al. provides further evidence that RST contributes to microwave noise.


A relevant question is whether the noise mechanisms identified in this study occur in the operational conditions of HEMTs. In actual InP HEMTs, the electric field under the gate is inhomogeneous and can be more than 20 kV/cm locally \cite{chen_effect_2021}. In this study, the highest electric field reached is only around 8 kV/cm. Further, the highest electric fields in HEMTs occur over tens of nanometers, while the minimum recess length in this study is 1 $\mu$m, and for these shorter recess lengths, RST was found to be less pronounced. 

The RST mechanism will play a role in HEMT noise if electrons are heated to high enough temperatures that they can thermionically emit into the barrier. It is well-known that electric fields in HEMTs occurring at the low-noise bias point can heat electrons significantly out of equilibrium with the lattice to temperatures exceeding 1000 K, even though the highest fields exist only over tens of nanometers. \cite{schwierz_modern_2003} If this were not the case, impact ionization and other high-bias effects would not occur. This temperature is hot enough to enable a fraction of a percent of channel electrons to thermionically emit into the barrier, producing a few hundred K of noise. \cite{esho_theory_2022} Therefore, this analysis and the findings of this study support the hypothesis that drain noise in HEMTs can be attributed to a combination of thermal and RST noise.




\section{Summary}

We have performed an experimental study of the microwave noise characteristics of InP TLM structures. The measured trends of noise for the two sample types are compatible with noise as arising from a combination of thermal and real-space transfer noise. This finding is expected to translate to the operational conditions of InP HEMTs. The contribution of RST to drain noise could potentially be mitigated by improving the quantum confinement of channel electrons, for instance by altering the alloy composition of the barrier. 


\begin{acknowledgments}

The authors thank Akim Babenko, Bekari Gabritchidze, Jan Grahn, Pekka Kangaslahti, Jacob Kooi, and Junjie Li for useful discussions. This work was sponsored by the Keck Institute for Space Studies and the National Science Foundation under  Grant No.\ 2511983. We gratefully acknowledge the critical support and infrastructure provided for this work by The Kavli Nanoscience Institute at the California Institute of Technology. 


\end{acknowledgments}

\bibliographystyle{aip}
\bibliography{Measurement}

@misc{Ardizzi_Variable_2025,
      title={Variable-temperature attenuator calibration method for on-wafer microwave noise characterization of low-noise amplifiers}, 
      author={Anthony J. Ardizzi and Jiayin Zhang and Akim A. Babenko and Kieran A. Cleary and Austin J. Minnich},
      year={2025},
      eprint={2511.13619},
      archivePrefix={arXiv},
      primaryClass={physics.ins-det},
      url={https://arxiv.org/abs/2511.13619}, 
}

@inproceedings{yuce_large-signal_2024,
	address = {Fort Lauderdale, FL, USA},
	title = {Large-{Signal} {Modeling} of a 50 nm {mHEMT} {Incorporating} a {Physical} {Impact}-{Ionization} {Model}},
	copyright = {https://doi.org/10.15223/policy-029},
	isbn = {979-8-3315-4124-8},
	url = {https://ieeexplore.ieee.org/document/10745715/},
	doi = {10.1109/BCICTS59662.2024.10745715},
	urldate = {2025-10-27},
	booktitle = {2024 {IEEE} {BiCMOS} and {Compound} {Semiconductor} {Integrated} {Circuits} and {Technology} {Symposium} ({BCICTS})},
	publisher = {IEEE},
	author = {Yüce, Yasin and Albahrani, Sayed Ali and Schwantuschke, Dirk and Moulin, Maxime and Leuther, Arnulf},
	month = oct,
	year = {2024},
	pages = {42--45},
}

@inproceedings{ruiz_impact_2019,
	address = {San Francisco, CA, USA},
	title = {Impact {Ionization} {Control} in 50 nm {Low}-{Noise} {High}-{Speed} {InP} {HEMTs} with {InAs} {Channel} {Insets}},
	copyright = {https://doi.org/10.15223/policy-029},
	isbn = {978-1-7281-4032-2},
	url = {https://ieeexplore.ieee.org/document/8993654/},
	doi = {10.1109/IEDM19573.2019.8993654},
	urldate = {2025-10-27},
	booktitle = {2019 {IEEE} {International} {Electron} {Devices} {Meeting} ({IEDM})},
	publisher = {IEEE},
	author = {Ruiz, Diego C. and Saranovac, Tamara and Han, Daxin and Ostinelli, Olivier and Bolognesi, C.R.},
	month = dec,
	year = {2019},
	pages = {9.3.1--9.3.4},
}

@article{cuadrado-calle_broadband_2017,
	title = {Broadband {MMIC} {LNAs} for {ALMA} {Band} 2+3 {With} {Noise} {Temperature} {Below} 28 {K}},
	volume = {65},
	copyright = {https://creativecommons.org/licenses/by/3.0/legalcode},
	issn = {0018-9480, 1557-9670},
	url = {https://ieeexplore.ieee.org/document/7836302/},
	doi = {10.1109/TMTT.2016.2639018},
	number = {5},
	urldate = {2025-09-06},
	journal = {IEEE Transactions on Microwave Theory and Techniques},
	author = {Cuadrado-Calle, David and George, Danielle and Fuller, Gary A. and Cleary, Kieran and Samoska, Lorene and Kangaslahti, Pekka and Kooi, Jacob W. and Soria, Mary and Varonen, Mikko and Lai, Richard and Mei, Xiaobing},
	month = may,
	year = {2017},
	pages = {1589--1597},
}

@article{heinz_50-nm_2021,
	title = {A 50-nm {Gate}-{Length} {Metamorphic} {HEMT} {Technology} {Optimized} for {Cryogenic} {Ultra}-{Low}-{Noise} {Operation}},
	volume = {69},
	copyright = {https://creativecommons.org/licenses/by/4.0/legalcode},
	issn = {0018-9480, 1557-9670},
	url = {https://ieeexplore.ieee.org/document/9459460/},
	doi = {10.1109/TMTT.2021.3081710},
	number = {8},
	urldate = {2025-09-06},
	journal = {IEEE Transactions on Microwave Theory and Techniques},
	author = {Heinz, Felix and Thome, Fabian and Leuther, Arnulf and Ambacher, Oliver},
	month = aug,
	year = {2021},
	pages = {3896--3907},
}

@article{aninkevicius_comparative_1993,
	title = {Comparative analysis of microwave noise in {GaAs} and {AlGaAs}/{GaAs} channels},
	volume = {36},
	copyright = {https://www.elsevier.com/tdm/userlicense/1.0/},
	issn = {00381101},
	url = {https://linkinghub.elsevier.com/retrieve/pii/003811019390174O},
	doi = {10.1016/0038-1101(93)90174-O},
	language = {en},
	number = {9},
	urldate = {2025-08-20},
	journal = {Solid-State Electronics},
	author = {Aninkevičius, V. and Bareikis, V. and Liberis, J. and Matulionis, A. and Sakalas, P.},
	month = sep,
	year = {1993},
	pages = {1339--1343},
}

@article{hess_negative_1979,
	title = {Negative differential resistance through real-space electron transfer},
	volume = {35},
	issn = {0003-6951, 1077-3118},
	url = {https://pubs.aip.org/apl/article/35/6/469/46110/Negative-differential-resistance-through-real},
	doi = {10.1063/1.91172},
	language = {en},
	number = {6},
	urldate = {2025-08-20},
	journal = {Applied Physics Letters},
	author = {Hess, K. and Morkoç, H. and Shichijo, H. and Streetman, B. G.},
	month = sep,
	year = {1979},
	pages = {469--471},
}

@article{gonzalez_noise_1997,
	title = {Noise suppression due to long-range {Coulomb} interaction: crossover between diffusive and ballistic transport regimes},
	volume = {12},
	issn = {0268-1242, 1361-6641},
	shorttitle = {Noise suppression due to long-range {Coulomb} interaction},
	url = {https://iopscience.iop.org/article/10.1088/0268-1242/12/8/001},
	doi = {10.1088/0268-1242/12/8/001},
	number = {8},
	urldate = {2025-08-20},
	journal = {Semiconductor Science and Technology},
	author = {González, T and Bulashenko, O M and Mateos, J and Pardo, D and Reggiani, L and Rubí, J M},
	month = aug,
	year = {1997},
	pages = {1053--1056},
}

@incollection{pucel_signal_1975,
	title = {Signal and {Noise} {Properties} of {Gallium} {Arsenide} {Microwave} {Field}-{Effect} {Transistors}},
	volume = {38},
	isbn = {978-0-12-014538-6},
	url = {https://linkinghub.elsevier.com/retrieve/pii/S0065253908612056},
	language = {en},
	urldate = {2025-08-20},
	booktitle = {Advances in {Electronics} and {Electron} {Physics}},
	publisher = {Elsevier},
	author = {Pucel, Robert A. and Haus, Hermann A. and Statz, Hermann},
	year = {1975},
	doi = {10.1016/S0065-2539(08)61205-6},
	pages = {195--265},
}

@article{pospieszalski_extremely_2005,
	title = {Extremely low-noise amplification with cryogenic {FETs} and {HFETs}: 1970-2004},
	volume = {6},
	copyright = {https://ieeexplore.ieee.org/Xplorehelp/downloads/license-information/IEEE.html},
	issn = {1527-3342},
	shorttitle = {Extremely low-noise amplification with cryogenic {FETs} and {HFETs}},
	url = {http://ieeexplore.ieee.org/document/1511915/},
	doi = {10.1109/MMW.2005.1511915},
	number = {3},
	urldate = {2025-08-07},
	journal = {IEEE Microwave Magazine},
	author = {Pospieszalski, M.W.},
	month = sep,
	year = {2005},
	pages = {62--75},
}

@article{bardin_microwaves_2021,
	title = {Microwaves in {Quantum} {Computing}},
	volume = {1},
	copyright = {https://creativecommons.org/licenses/by/4.0/legalcode},
	issn = {2692-8388},
	url = {https://ieeexplore.ieee.org/document/9318753/},
	doi = {10.1109/JMW.2020.3034071},
	number = {1},
	urldate = {2025-08-07},
	journal = {IEEE Journal of Microwaves},
	author = {Bardin, Joseph C. and Slichter, Daniel H. and Reilly, David J.},
	month = jan,
	year = {2021},
	pages = {403--427},
}

@article{chen_effect_2021,
	title = {Effect of {Fixed} {Charges} at {Interface} {Between} {InP} and {Bonding} {Layer} on {Heterogeneous} {Integration} of {InP} {HEMTs}},
	volume = {68},
	copyright = {https://ieeexplore.ieee.org/Xplorehelp/downloads/license-information/IEEE.html},
	issn = {0018-9383, 1557-9646},
	url = {https://ieeexplore.ieee.org/document/9385930/},
	doi = {10.1109/TED.2021.3066139},
	number = {5},
	urldate = {2025-07-28},
	journal = {IEEE Transactions on Electron Devices},
	author = {Chen, Yao and Yang, Lin-An and Jin, Zhi and Su, Yong-Bo and Hao, Yue},
	month = may,
	year = {2021},
	pages = {2226--2232},
}

@article{lee_theoretical_2021,
	title = {Theoretical and experimental analysis of the source resistance components in {In0}.{7Ga0}.{3As} quantum-well high-electron-mobility transistors},
	volume = {78},
	copyright = {http://www.springer.com/tdm},
	issn = {0374-4884, 1976-8524},
	url = {http://link.springer.com/10.1007/s40042-021-00096-0},
	doi = {10.1007/s40042-021-00096-0},
	language = {en},
	number = {6},
	urldate = {2025-07-11},
	journal = {Journal of the Korean Physical Society},
	author = {Lee, In-Geun and Ko, Dae-Hong and Yun, Seung-Won and Kim, Jun-Gyu and Jo, Hyeon-Bhin and Kim, Dae-Hyun and Tsutsumi, Takuya and Sugiyama, Hiroki and Matsuzaki, Hideaki},
	month = mar,
	year = {2021},
	note = {Publisher: Springer Science and Business Media LLC},
	pages = {516--522},
}

@inproceedings{pospieszalski_dependence_2017,
	address = {Phoenix, AZ, USA},
	title = {On the dependence of {FET} noise model parameters on ambient temperature},
	isbn = {978-1-5090-3446-8},
	url = {http://ieeexplore.ieee.org/document/7885975/},
	doi = {10.1109/RWS.2017.7885975},
	urldate = {2025-06-12},
	booktitle = {2017 {IEEE} {Radio} and {Wireless} {Symposium} ({RWS})},
	publisher = {IEEE},
	author = {Pospieszalski, Marian W.},
	month = jan,
	year = {2017},
	pages = {159--161},
}

@article{shigekawa_impact-ionization-induced_1999,
	title = {Impact-ionization-induced noise in {InGaAs}-based 0.1-μm-gate {HEMTs}},
	volume = {272},
	copyright = {https://www.elsevier.com/tdm/userlicense/1.0/},
	issn = {09214526},
	url = {https://linkinghub.elsevier.com/retrieve/pii/S0921452699003397},
	doi = {10.1016/S0921-4526(99)00339-7},
	number = {1-4},
	urldate = {2025-06-11},
	journal = {Physica B: Condensed Matter},
	author = {Shigekawa, N},
	month = dec,
	year = {1999},
	pages = {562--564},
}

@article{shah_investigation_1981,
	title = {{INVESTIGATION} {OF} {HOT} {CARRIER} {RELAXATION} {WITH} {PICOSECOND} {LASER} {PULSES}},
	volume = {42},
	issn = {0449-1947},
	url = {http://www.edpsciences.org/10.1051/jphyscol:1981755},
	doi = {10.1051/jphyscol:1981755},
	number = {C7},
	urldate = {2025-06-11},
	journal = {Le Journal de Physique Colloques},
	author = {Shah, J.},
	month = oct,
	year = {1981},
	pages = {C7--445--C7--462},
}

@article{matulionis_hot-electron_1999,
	title = {Hot-electron energy relaxation, noise, and lattice strain in {InGaAs} quantum well channels},
	volume = {74},
	issn = {0003-6951, 1077-3118},
	url = {https://pubs.aip.org/apl/article/74/13/1895/70002/Hot-electron-energy-relaxation-noise-and-lattice},
	doi = {10.1063/1.123705},
	language = {en},
	number = {13},
	urldate = {2025-06-11},
	journal = {Applied Physics Letters},
	author = {Matulionis, A. and Aninkevičius, V. and Liberis, J. and Matulionienė, I. and Berntgen, J. and Heime, K. and Hartnagel, H. L.},
	month = mar,
	year = {1999},
	pages = {1895--1897},
}

@article{matulionis_qw-shape-dependent_1997,
	title = {{QW}-{Shape}-{Dependent} {Hot}-{Electron} {Velocity} {Fluctuations} in {InGaAs}-{Based} {Heterostructures}},
	volume = {204},
	copyright = {http://doi.wiley.com/10.1002/tdm\_license\_1.1},
	issn = {03701972, 15213951},
	url = {https://onlinelibrary.wiley.com/doi/10.1002/1521-3951(199711)204:1<453::AID-PSSB453>3.0.CO;2-V},
	doi = {10.1002/1521-3951(199711)204:1<453::AID-PSSB453>3.0.CO;2-V},
	language = {en},
	number = {1},
	urldate = {2025-06-11},
	journal = {physica status solidi (b)},
	author = {Matulionis, A. and Aninkevičius, V. and Berntgen, J. and Gasquet, D. and Liberis, J. and Matulionienė, I.},
	month = nov,
	year = {1997},
	pages = {453--455},
}

@article{cha_optimization_2023,
	title = {Optimization of {Channel} {Structures} in {InP} {HEMT} {Technology} for {Cryogenic} {Low}-{Noise} and {Low}-{Power} {Operation}},
	volume = {70},
	copyright = {https://ieeexplore.ieee.org/Xplorehelp/downloads/license-information/IEEE.html},
	issn = {0018-9383, 1557-9646},
	url = {https://ieeexplore.ieee.org/document/10070577/},
	doi = {10.1109/TED.2023.3255160},
	number = {5},
	urldate = {2025-06-11},
	journal = {IEEE Transactions on Electron Devices},
	author = {Cha, Eunjung and Wadefalk, Niklas and Moschetti, Giuseppe and Pourkabirian, Arsalan and Stenarson, Jörgen and Li, Junjie and Kim, Dae-Hyun and Grahn, Jan},
	month = may,
	year = {2023},
	pages = {2431--2436},
}

@article{li_investigation_2024,
	title = {Investigation of {Noise} {Properties} in the {InP} {HEMT} for {LNAs} in {Qubit} {Amplification}: {Effects} {From} {Channel} {Indium} {Content}},
	volume = {12},
	copyright = {https://creativecommons.org/licenses/by-nc-nd/4.0/},
	issn = {2168-6734},
	shorttitle = {Investigation of {Noise} {Properties} in the {InP} {HEMT} for {LNAs} in {Qubit} {Amplification}},
	url = {https://ieeexplore.ieee.org/document/10454583/},
	doi = {10.1109/JEDS.2024.3371905},
	urldate = {2025-06-11},
	journal = {IEEE Journal of the Electron Devices Society},
	author = {Li, Junjie and Bergsten, Johan and Pourkabirian, Arsalan and Grahn, Jan},
	year = {2024},
	pages = {243--248},
}

@article{esho_theory_2022,
	title = {Theory of drain noise in high electron mobility transistors based on real-space transfer},
	volume = {131},
	issn = {0021-8979, 1089-7550},
	url = {https://pubs.aip.org/jap/article/131/8/085111/2836567/Theory-of-drain-noise-in-high-electron-mobility},
	doi = {10.1063/5.0069352},
	language = {en},
	number = {8},
	urldate = {2025-06-11},
	journal = {Journal of Applied Physics},
	author = {Esho, Iretomiwa and Choi, Alexander Y. and Minnich, Austin J.},
	month = feb,
	year = {2022},
	pages = {085111},
}

@inproceedings{gabritchidze_experimental_2022,
	address = {Denver, CO, USA},
	title = {Experimental {Characterization} of {Temperature}-{Dependent} {Microwave} {Noise} of {Discrete} {HEMTs}: {Drain} {Noise} and {Real}-{Space} {Transfer}},
	copyright = {https://doi.org/10.15223/policy-029},
	isbn = {978-1-66549-613-1},
	shorttitle = {Experimental {Characterization} of {Temperature}-{Dependent} {Microwave} {Noise} of {Discrete} {HEMTs}},
	url = {https://ieeexplore.ieee.org/document/9865505/},
	doi = {10.1109/IMS37962.2022.9865505},
	urldate = {2025-06-11},
	booktitle = {2022 {IEEE}/{MTT}-{S} {International} {Microwave} {Symposium} - {IMS} 2022},
	publisher = {IEEE},
	author = {Gabritchidze, Bekari and Cleary, Kieran and Kooi, Jacob and Esho, Iretomiwa and Readhead, Anthony C. and Minnich, Austin J.},
	month = jun,
	year = {2022},
	pages = {615--618},
}

@article{luryi_charge_1984,
	title = {Charge injection transistor based on real-space hot-electron transfer},
	volume = {31},
	issn = {1557-9646},
	url = {https://ieeexplore.ieee.org/document/1483901/},
	doi = {10.1109/T-ED.1984.21616},
	number = {6},
	urldate = {2025-11-05},
	journal = {IEEE Transactions on Electron Devices},
	author = {Luryi, S. and Kastalsky, A. and Gossard, A.C. and Hendel, R.H.},
	month = jun,
	year = {1984},
	pages = {832--839},
}

@article{kastalsky_novel_1983,
	title = {Novel real-space hot-electron transfer devices},
	volume = {4},
	issn = {1558-0563},
	url = {https://ieeexplore.ieee.org/document/1483497/},
	doi = {10.1109/EDL.1983.25753},
	number = {9},
	urldate = {2025-11-05},
	journal = {IEEE Electron Device Letters},
	author = {Kastalsky, A. and Luryi, S.},
	month = sep,
	year = {1983},
	pages = {334--336},
}

@article{price_intervalley_1960,
	title = {Intervalley {Noise}},
	volume = {31},
	issn = {0021-8979, 1089-7550},
	url = {https://pubs.aip.org/jap/article/31/6/949/163184/Intervalley-Noise},
	doi = {10.1063/1.1735782},
	language = {en},
	number = {6},
	urldate = {2025-11-05},
	journal = {Journal of Applied Physics},
	author = {Price, P. J.},
	month = jun,
	year = {1960},
	pages = {949--953},
}

@techreport{arthur_measurement_1974,
	address = {Gaithersburg, MD},
	title = {The {Measurement} of noise performance factors : a metrology guide},
	shorttitle = {The {Measurement} of noise performance factors},
	url = {https://nvlpubs.nist.gov/nistpubs/Legacy/MONO/nbsmonograph142.pdf},
	language = {en},
	number = {NBS MONO 142},
	urldate = {2025-10-28},
	institution = {National Bureau of Standards},
	author = {Arthur, M G and Anson, W J},
	year = {1974},
	doi = {10.6028/NBS.MONO.142},
	note = {Edition: 0},
	pages = {NBS MONO 142},
}

@article{skullerud_longitudinal_1969,
	title = {Longitudinal diffusion of electrons in electrostatic fields in gases},
	volume = {2},
	issn = {00223700},
	url = {https://iopscience.iop.org/article/10.1088/0022-3700/2/6/309},
	doi = {10.1088/0022-3700/2/6/309},
	number = {6},
	urldate = {2025-08-27},
	journal = {Journal of Physics B: Atomic and Molecular Physics},
	author = {Skullerud, H R},
	month = jun,
	year = {1969},
	pages = {696--705},
}

@article{gantsevich_theory_1979,
	title = {Theory of fluctuations in nonequilibrium electron gas},
	volume = {2},
	copyright = {http://www.springer.com/tdm},
	issn = {1826-9850},
	url = {http://link.springer.com/10.1007/BF02724353},
	doi = {10.1007/BF02724353},
	language = {en},
	number = {5},
	urldate = {2025-08-27},
	journal = {La Rivista del Nuovo Cimento},
	author = {Gantsevich, S. V. and Gurevich, V. L. and Katilius, R.},
	month = may,
	year = {1979},
	pages = {1--88},
}

@book{hartnagel_microwave_2001,
	address = {New York},
	title = {Microwave noise in semiconductor devices},
	isbn = {978-0-471-38432-8},
	publisher = {Wiley},
	author = {Hartnagel, Hans and Katilius, Ramunas and Matulionis, A.},
	year = {2001},
}

@phdthesis{li_epitaxial_2024,
	address = {Terahetz and Millimetre Wave Laboratory, Department of Microtechnology and Nanoscience, Chalmers University of Technology, Gothenburg, Sweden},
	title = {Epitaxial {Optimization} of {InP} {High} {Electron} {Mobility} {Transistors} in {Low}-{Noise} {Amplifiers} for {Qubit} {Readout}},
	language = {en},
	school = {Chalmers University of Technology},
	author = {Li, Junjie},
	year = {2024},
}

@inproceedings{jo_l_2020,
	address = {San Francisco, CA, USA},
	title = {L $_{\textrm{g}}$ = 19 nm {In} $_{\textrm{0.8}}$ {Ga} $_{\textrm{0.2}}$ {As} composite-channel {HEMTs} with f $_{\textrm{{T}}}$ = 738 {GHz} and f $_{\textrm{max}}$ = 492 {GHz}},
	copyright = {https://ieeexplore.ieee.org/Xplorehelp/downloads/license-information/IEEE.html},
	isbn = {978-1-72818-888-1},
	url = {https://ieeexplore.ieee.org/document/9372070/},
	doi = {10.1109/IEDM13553.2020.9372070},
	language = {en},
	urldate = {2024-03-27},
	booktitle = {2020 {IEEE} {International} {Electron} {Devices} {Meeting} ({IEDM})},
	publisher = {IEEE},
	author = {Jo, Hyeon-Bhin and Yun, Seung-Won and Kim, Jun-Gyu and Yun, Do-Young and Lee, In-Geun and Kim, Dae-Hyun and Kim, Tae-Woo and Kim, Sang-Kuk and Yun, Jacob and Kim, Ted and Tsutsumi, Takuya and Sugiyama, Hiroki and Matsuzaki, Hideaki},
	month = dec,
	year = {2020},
	pages = {8.4.1--8.4.4},
}

@book{ambrozy_electronic_1982,
	address = {New York ; London},
	title = {Electronic noise},
	isbn = {978-0-07-001124-3},
	language = {enghun},
	publisher = {McGraw-Hill International Book Co},
	author = {Ambrózy, András},
	year = {1982},
}

@article{vurgaftman_band_2001,
	title = {Band parameters for {III}-{V} compound semiconductors and their alloys},
	volume = {89},
	issn = {0021-8979, 1089-7550},
	url = {http://aip.scitation.org/doi/10.1063/1.1368156},
	doi = {10.1063/1.1368156},
	language = {en},
	number = {11},
	urldate = {2022-12-07},
	journal = {Journal of Applied Physics},
	author = {Vurgaftman, I. and Meyer, J. R. and Ram-Mohan, L. R.},
	month = jun,
	year = {2001},
	pages = {5815--5875},
}

@book{schwierz_modern_2003,
	address = {New York},
	title = {Modern microwave transistors: theory, design, and performance},
	isbn = {978-0-471-41778-1},
	shorttitle = {Modern microwave transistors},
	publisher = {Wiley-Interscience},
	author = {Schwierz, Frank and Liou, Juin J.},
	year = {2003},
}

@phdthesis{ardizzi_anthony_j_self-heating_2022,
	title = {Self-{Heating} of {HEMT} {Low}-{Noise} {Amplifiers} in {Liquid} {Cryogenic} {Environments} and the {Limits} of {Microwave} {Noise} {Performance}},
	copyright = {No commercial reproduction, distribution, display or performance rights in this work are provided.},
	url = {https://resolver.caltech.edu/CaltechTHESIS:05272022-180844058},
	language = {en},
	urldate = {2023-02-27},
	school = {California Institute of Technology},
	author = {Ardizzi, Anthony J.},
	month = jun,
	year = {2022},
	doi = {10.7907/9VA8-TC64},
	note = {Medium: PDF
Version Number: Final},
}

@article{russell_cryogenic_2012,
	title = {Cryogenic {Self}-{Calibrating} {Noise} {Parameter} {Measurement} {System}},
	volume = {60},
	issn = {1557-9670},
	doi = {10.1109/TMTT.2012.2188813},
	number = {5},
	journal = {IEEE Transactions on Microwave Theory and Techniques},
	author = {Russell, Damon and Weinreb, Sander},
	month = may,
	year = {2012},
	note = {Conference Name: IEEE Transactions on Microwave Theory and Techniques},
	pages = {1456--1467},
}

@phdthesis{cha_inp_2020,
	address = {Göteborg},
	title = {{InP} high electron mobility transistors for cryogenic low noise and low power amplifiers},
	language = {English},
	school = {Chalmers Univ. of Technology},
	author = {Cha, Eunjung},
	year = {2020},
}

@inproceedings{lamarre_ohmic_1993,
	title = {Ohmic contacts with different metal structures for lattice matched {InP} based heterostructures},
	doi = {10.1109/ICIPRM.1993.380642},
	booktitle = {1993 (5th) {International} {Conference} on {Indium} {Phosphide} and {Related} {Materials}},
	author = {Lamarre, P. and McTaggart, R. and Pulley, M. and Huang, J. and Jackson, G.},
	month = apr,
	year = {1993},
	pages = {333--336},
}

@article{malmkvist_epitaxial_2008,
	title = {Epitaxial {Optimization} of 130-nm {Gate}-{Length} {InGaAs}/{InAlAs}/{InP} {HEMTs} for {High}-{Frequency} {Applications}},
	volume = {55},
	issn = {1557-9646},
	doi = {10.1109/TED.2007.910613},
	number = {1},
	journal = {IEEE Transactions on Electron Devices},
	author = {Malmkvist, Mikael and Wang, Shumin and Grahn, Jan V.},
	month = jan,
	year = {2008},
	note = {Conference Name: IEEE Transactions on Electron Devices},
	pages = {268--275},
}

@article{ardizzi_self-heating_2022,
	title = {Self-heating of cryogenic high electron-mobility transistor amplifiers and the limits of microwave noise performance},
	volume = {132},
	copyright = {All rights reserved},
	issn = {0021-8979, 1089-7550},
	url = {https://aip.scitation.org/doi/10.1063/5.0103156},
	doi = {10.1063/5.0103156},
	language = {en},
	number = {8},
	urldate = {2022-08-30},
	journal = {Journal of Applied Physics},
	author = {Ardizzi, Anthony J. and Choi, Alexander Y. and Gabritchidze, Bekari and Kooi, Jacob and Cleary, Kieran A. and Readhead, Anthony C. and Minnich, Austin J.},
	month = aug,
	year = {2022},
	pages = {084501},
}

@inproceedings{pospieszalski_limits_2017,
	title = {On the limits of noise performance of field effect transistors},
	doi = {10.1109/MWSYM.2017.8059045},
	booktitle = {2017 {IEEE} {MTT}-{S} {International} {Microwave} {Symposium} ({IMS})},
	publisher = {IEEE},
	author = {Pospieszalski, Marian W.},
	month = jun,
	year = {2017},
	pages = {1953--1956},
}

@book{sze_physics_2007,
	address = {Hoboken, N.J},
	edition = {3rd ed},
	title = {Physics of semiconductor devices},
	isbn = {978-0-471-14323-9},
	publisher = {Wiley-Interscience},
	author = {Sze, S. M. and Ng, Kwok Kwok},
	year = {2007},
}

@article{krantz_quantum_2019,
	title = {A quantum engineer's guide to superconducting qubits},
	volume = {6},
	issn = {1931-9401},
	url = {http://aip.scitation.org/doi/10.1063/1.5089550},
	doi = {10.1063/1.5089550},
	language = {en},
	number = {2},
	urldate = {2022-03-30},
	journal = {Applied Physics Reviews},
	author = {Krantz, P. and Kjaergaard, M. and Yan, F. and Orlando, T. P. and Gustavsson, S. and Oliver, W. D.},
	month = jun,
	year = {2019},
	pages = {021318},
}

@article{schleeh_characterization_2013,
	title = {Characterization and {Modeling} of {Cryogenic} {Ultralow}-{Noise} {InP} {HEMTs}},
	volume = {60},
	issn = {1557-9646},
	doi = {10.1109/TED.2012.2227485},
	number = {1},
	journal = {IEEE Transactions on Electron Devices},
	author = {Schleeh, Joel and Rodilla, Helena and Wadefalk, Niklas and Nilsson, Per-Ake and Grahn, Jan},
	month = jan,
	year = {2013},
	note = {Conference Name: IEEE Transactions on Electron Devices},
	pages = {206--212},
}

@article{pospieszalski_modeling_1989,
	title = {Modeling of noise parameters of {MESFETs} and {MODFETs} and their frequency and temperature dependence},
	volume = {37},
	issn = {1557-9670},
	doi = {10.1109/22.32217},
	number = {9},
	journal = {IEEE Transactions on Microwave Theory and Techniques},
	author = {Pospieszalski, M. W.},
	month = sep,
	year = {1989},
	note = {Conference Name: IEEE Transactions on Microwave Theory and Techniques},
	pages = {1340--1350},
}

@article{schleeh_phonon_2015,
	title = {Phonon black-body radiation limit for heat dissipation in electronics},
	volume = {14},
	issn = {1476-1122, 1476-4660},
	url = {http://www.nature.com/articles/nmat4126},
	doi = {10.1038/nmat4126},
	language = {en},
	number = {2},
	urldate = {2021-03-22},
	journal = {Nature Materials},
	author = {Schleeh, J. and Mateos, J. and Íñiguez-de-la-Torre, I. and Wadefalk, N. and Nilsson, P. A. and Grahn, J. and Minnich, A. J.},
	month = feb,
	year = {2015},
	pages = {187--192},
}

\end{document}